\documentclass[12pt]{article}

\usepackage{graphicx}
\usepackage{epsf}

\usepackage{amsmath,amssymb,float} 

\usepackage{pstricks}
\usepackage{color}

\def\beq{\begin{equation}}
\def\eq{\end{equation}}
\def\eeq{\end{equation}}

\def\centeron#1#2{{\setbox0=\hbox{#1}\setbox1=\hbox{#2}\ifdim
\wd1>\wd0\kern.5\wd1\kern-.5\wd0\fi
\copy0\kern-.5\wd0\kern-.5\wd1\copy1\ifdim\wd0>\wd1
\kern.5\wd0\kern-.5\wd1\fi}}
\def\ltap{\;\centeron{\raise.35ex\hbox{$<$}}{\lower.65ex\hbox{$\sim$}}\;}
\def\gtap{\;\centeron{\raise.35ex\hbox{$>$}}{\lower.65ex\hbox{$\sim$}}\;}

\def\MET{{\not \!  \! E}_T}

\def\chii0{\chi_i^0}
\def\chij0{\chi_j^0}

\def\GTLT{ \mathop{}_{<}^{>} }

\def\foursqr#1#2{{\vcenter{\vbox{
 \hrule height.#2pt
 \hbox{\vrule width.#2pt height#1pt \kern#1pt
 \vrule width.#2pt}
 \hrule height.#2pt
 \hrule height.#2pt
 \hbox{\vrule width.#2pt height#1pt \kern#1pt
 \vrule width.#2pt}
 \hrule height.#2pt
     \hrule height.#2pt
 \hbox{\vrule width.#2pt height#1pt \kern#1pt
 \vrule width.#2pt}
 \hrule height.#2pt
     \hrule height.#2pt
 \hbox{\vrule width.#2pt height#1pt \kern#1pt
 \vrule width.#2pt}
 \hrule height.#2pt}}}}
\def\psqr#1#2{{\vcenter{\vbox{\hrule height.#2pt
 \hbox{\vrule width.#2pt height#1pt \kern#1pt
 \vrule width.#2pt}
 \hrule height.#2pt \hrule height.#2pt
 \hbox{\vrule width.#2pt height#1pt \kern#1pt
 \vrule width.#2pt}
 \hrule height.#2pt}}}}
\def\sqr#1#2{{\vcenter{\vbox{\hrule height.#2pt
 \hbox{\vrule width.#2pt height#1pt \kern#1pt
 \vrule width.#2pt}
 \hrule height.#2pt}}}}

\def\figin{\epsfcheck\figin}\def\figins{\epsfcheck\figins}
\def\epsfcheck{\ifx\epsfbox\UnDeFiNeD
\message{(NO epsf.tex, FIGURES WILL BE IGNORED)}
\gdef\figin##1{\vskip2in}\gdef\figins##1{\hskip.5in}
\else\message{(FIGURES WILL BE INCLUDED)}%
\gdef\figin##1{##1}\gdef\figins##1{##1}\fi}
\def\DefWarn#1{}
\def\figinsert{\goodbreak\midinsert}
\def\ifig#1#2#3{\DefWarn#1\xdef#1{fig.~\the\figno}
\writedef{#1\leftbracket fig.\noexpand~\the\figno}%
\figinsert\figin{\centerline{#3}}\medskip\centerline{\vbox{\baselineskip12pt
\advance\hsize by -1truein\noindent\footnotefont{\bf
Fig.~\the\figno:\ } \it#2}}
\bigskip\endinsert\global\advance\figno by1}

\def\fig#1#2#3#4{\vskip 0.5cm \begingroup \midinsert \centerline{
\psfig{file=#1,width=#2}} \vskip 0.4cm
\global\advance\figno by 1
\centerline{\vbox{\baselineskip=12pt \noindent Figure \the\figno: #3}}
\endinsert \endgroup {\xdef#4{\the\figno}} }

\def\figcrop#1#2#3#4#5#6#7#8{\vskip 0.5cm \begingroup \midinsert \centerline{
\psfig{file=#1,width=#2,bbllx=#3,bblly=#4,bburx=#5,bbury=#6}} \vskip 0.4cm
\global\advance\figno by 1
\centerline{\vbox{\baselineskip=12pt \noindent Figure \the\figno: #7}}
\endinsert \endgroup {\xdef#8{\the\figno}} \vskip .5cm}

\def\figlabel#1{\xdef#1{\the\figno}}
\def\encadremath#1{\vbox{\hrule\hbox{\vrule\kern8pt\vbox{\kern8pt
\hbox{$\displaystyle #1$}\kern8pt}
\kern8pt\vrule}\hrule}}
\def\underarrow#1{\vbox{\ialign{##\crcr$\hfil\displaystyle
 {#1}\hfil$\crcr\noalign{\kern1pt\nointerlineskip}$\longrightarrow$\crcr}}}

\begin{document}

\begin{titlepage}

\begin{center}
\vspace*{-1cm}

\hfill RU-NHETC-2011-21 \\
\hfill UTTG-25-11 \\
\hfill TCC-028-11 \\
\vskip 0.65in
{\LARGE \bf Multi-Lepton Signals of the Higgs Boson} \\
\vspace{.15in}

\vskip 0.35in
{\large Emmanuel Contreras-Campana},$^1$~
{\large Nathaniel Craig},$^{1,2}$
\vskip0.1in
{\large Richard Gray},$^1$~
{\large Can Kilic},$^3$~
{\large  Michael Park},$^1$
\vskip0.1in
{\large Sunil Somalwar},$^1$~
{\rm and}~
{\large Scott Thomas}$^1$

\vskip 0.25in
$^1${\em 
Department of Physics \\
Rutgers University \\
Piscataway, NJ 08854}

\vskip 0.12in
$^2${\em 
School of Natural Sciences
\\ Institute for Advanced Study \\
Princeton, NJ 08540 \\}

\vskip 0.12in
$^3${\em
Theory Group, Department of Physics and Texas Cosmology Center \\
The University of Texas at Austin \\
Austin, TX 78712}

\vskip 0.4in

\end{center}

\baselineskip=16pt

\begin{abstract}

\noindent

The possibility of searching for the Higgs boson in channels with multiple non-resonant leptons is evaluated in light of recent advances in multi-lepton search techniques at the LHC. The total multi-lepton Higgs signal exceeds the four lepton gold-plated resonant mode, but is spread over many channels with same-sign di-lepton, tri-lepton, and four lepton final states. While any individual channel alone is not significant, the exclusive combination across multiple channels is shown to provide a sensitivity competitive with other discovery level searches for the Higgs boson. We estimate that with 5 fb$^{-1}$ of data, existing non-optimized multi-lepton searches at the LHC could exclude the Higgs boson to 95\% CL at a few times the predicted Standard Model cross section in the mass range $120-150$ GeV. Refinements focused specifically on the Higgs boson signal are suggested that would further increase sensitivity. We illustrate the possibility of discerning patterns in production and decay modes using correlations across multiple channels by comparing sensitivities to Standard Model, Fermi-phobic, and $b$-phobic Higgs bosons.

\end{abstract}

\end{titlepage}

\baselineskip=17pt

\newpage





\section{Introduction}

The discovery and characterization of the Higgs boson are among the central aims of the physics program at the Large Hadron Collider (LHC). The well-defined production and decay modes of the Standard Model (SM) Higgs boson allow for mass-dependent searches tailored to a variety of specific channels (for a review, see \cite{PDG} and references therein). These searches grow increasingly compelling as various channels gain the sensitivity to discover or exclude the Standard Model Higgs across a wide range of masses. Current data from LEP, the Tevatron, and the LHC already constrain a light SM Higgs to lie within a narrowly proscribed region \cite{Higgsrefs}.

The dominant production channel of the SM Higgs at hadron colliders is through gluon-gluon fusion ($gg$F), $gg \to h$. Existing LHC Higgs searches are typically tailored towards this production channel due to both the large cross section and the resulting Higgs resonance. There are also a variety of ancillary channels in which the Higgs is produced in association with other quarks or vector bosons. These are, in order of decreasing production rate: weak vector boson fusion (VBF), $qq \to qqh$; $Wh$ and $Zh$ associated production (or {\it Higgs-strahlung}), $q\bar q' \to Wh, Zh$; and $t \bar t h$ associated production, $q\bar q, gg \to t \bar t h$. Loosely speaking, the cross sections for weak VBF and $Wh, Zh$ associated production are an order of magnitude smaller than that of $gg$F, while $t\bar th$ associated production is smaller by a further order of magnitude.  Di-Higgs production through gluon-gluon fusion, $g g\to hh,$ is smaller by roughly a further order of magnitude. Nonetheless, they may provide interesting alternative routes to the discovery of the Higgs.

Whatever the mechanism of Higgs production, current search strategies are principally governed by the decay products of the Higgs. The primary decay modes for a light Higgs include $h \to b \bar b,\, \tau^+ \tau^-, \, c \bar c,  \, gg,  \, WW^*, \, ZZ^*, \, \gamma \gamma,$ and $Z \gamma$. Branching ratios to these final states are a sensitive function of the Higgs mass, with $b \bar b, \, \tau^+ \tau^-,$ and $gg$ dominating at low masses ($m_h \lesssim 135$ GeV) and $WW^*, \, ZZ^*$ dominating at higher masses. The colored final states $b \bar b, c \bar c,$ and $gg$ are inauspicious search modes at the LHC due to large QCD backgrounds; more promising are the diphoton channel and the leptonic final states of the $WW^*, ZZ^*,$ and $\tau^+ \tau^-$ channels.

The production and decay modes of the Higgs lead to a variety of possible search strategies at hadron colliders. At the LHC, the three main search methods with the greatest discovery potential are $h \to \gamma \gamma$, $h \to ZZ^* \to 4 \ell$, and $h \to WW^* \to \ell \ell \nu \nu$. Although the branching ratio for $\gamma \gamma$ is small, the distinctive final state topology makes it a crucial search channel for lighter masses. At higher masses the increased branching fraction to $WW^*$ and $ZZ^*$, combined with the cleanliness of $2 \ell$ and $4 \ell$ final states, make $h \to \ell \ell \nu \nu$ and $h \to 4 \ell$ particularly attractive. Significantly, both $h \to ZZ^* \to 4 \ell$ and $h \to \gamma \gamma$ (the so-called gold- and silver-plated channels) are {\it resonant}  search modes, in that the invariant mass of the final state reconstructs the Higgs mass. This allows the direct determination of the Higgs mass, but at the expense of sensitivity to, e.g., nonresonant $4 \ell$ final states. This is in contrast to the $h \to WW^* \to \ell \ell \nu \nu$ channel, in which the missing energy from the neutrinos makes reconstructing the Higgs mass more challenging. Individually and in combination, these search channels are growing ever closer to constraining the production of a Standard Model Higgs boson in the light mass window, although backgrounds for these channels are large and potentially quite subtle.\footnote{For instance, asymmetric internal photon conversions as a source of fake leptons in Higgs searches has only recently been studied in detail~\cite{Gray:2011us}.}

However, despite the current focus on the ggF channel, there may also be considerable sensitivity to final states populated predominantly by {\it associated} production channels of the Higgs boson. Although these have weaker prospects for the determination of the Higgs mass based on kinematics, they often lead to final states with particularly low Standard Model backgrounds. In particular, the multitude of 3- and 4-lepton final states available from Higgs production in association with $W$ and $Z$ bosons or a $t \bar t$ pair provides a key handle on picking the Higgs signal out of Standard Model backgrounds. Although searches in some specific individual production and decay channels
have been proposed previously  \cite{Baer:1998cm}, recent advances in multi-lepton searches at the LHC \cite{CMSMulti} have brought the possibility of a dedicated multi-lepton Higgs search across
multiple channels into sharp focus.

Such a multi-lepton search enjoys several advantages. Standard Model backgrounds to multi-lepton processes are quite low, particularly in the absence of an on-shell $Z$ boson. Further discrimination may be obtained by looking in regions of high missing energy or hadronic activity, away from typical Standard Model processes. Ultimately, perhaps the greatest advantage lies in the plethora of possible multi-lepton channels; more sensitivity to  Higgs searches may be added by combining various $3\ell$ and $4 \ell$ channels (as well as same-sign $2 \ell$ channels), particularly those that do not reconstruct an on-shell $Z$ or the Higgs resonance itself.

In this paper we pursue a simple goal: applying the existing CMS multi-lepton search strategy to the Higgs boson in order to determine how effective a new low-background, multi-channel analysis may be in the hunt for the Higgs. To this end, we focus on the Higgs production and decay channels most likely to produce $3 \ell$ and $4 \ell$ final states. These are dominated by $Wh, Zh$ and $t \bar t h$ associated production with $h \to WW^*, ZZ^*$. At low masses, significant contributions may also arise from $h \to \tau^+ \tau^-$ with both $\tau$'s decaying leptonically. Additional contributions to $h \to 3 \ell$ and {\it nonresonant} $h \to 4 \ell$ arise from the dominant production modes, $gg$ fusion and vector boson fusion, where $h \to ZZ^* \to \ell \ell \tau \tau$ and the $\tau$'s decay leptonically. Finally, a surprisingly significant contribution to resonant multi-lepton final states not covered by current resonant searches arises at low mass (particularly $m_h \lesssim 130$ GeV) when the Higgs decays to two off-shell $Z$ bosons, $h \to Z^*Z^* \to 4 \ell$. Taken together, the signal of these multi-lepton modes exceeds that of the gold-plated resonant $4 \ell$ mode. Exploring the sensitivity of existing multi-lepton searches to these production channels may allow the development of a search tailored channel-by-channel toward the discovery and characterization of the Higgs.

The paper is organized as follows: In \S\ref{sec:search} we discuss the signal channels and sensitivity of multi-lepton searches to Higgs production, closely following the current multi-lepton search strategy of \cite{CMSMulti}. In \S\ref{sec:future} we explore potential improvements that may be made to tailor the multi-lepton search towards the Higgs. Concluding remarks are presented in \S\ref{sec:conc}.


\section{A multi-lepton search for the Higgs} \label{sec:search}

Multi-lepton searches have already been employed to good effect at the early LHC \cite{ CMSMulti}, as lepton-rich Standard Model processes are relatively rare. To date these searches have been primarily interpreted in terms of supersymmetry and other exotica; here we simply wish to probe the sensitivity of existing search strategies to the Standard Model Higgs and its variations.  Our objective is to determine how effective a new low-background, multi-channel analysis may be in the hunt for the Higgs. While simply applying the existing multi-lepton search strategy to a Higgs signal is not optimal ``out of the box'',  it illustrates the considerable power of combining several low-background channels in the same search. The sensitivity of each individual channel to Higgs production may subsequently be improved by tailoring cuts to the corresponding dominant production mode for that channel. Thus our analysis is, in part, intended as an exercise to determine how these individual channels might best be optimized for a dedicated Higgs search. While this section will focus on Higgs signals in the existing multi-lepton search, we will present suggestions for refinement in \S\ref{sec:future}.

\subsection{Multi-lepton signal channels}\label{subsec:channels}

In order to most directly probe the sensitivity of current multi-lepton searches to the Higgs boson, we will apply the specific preselection and analysis cuts detailed in \cite{CMSMulti} -- and, indeed, the broader search strategy --  to Higgs signal events.

The Standard Model background to multi-lepton searches is small and may be further reduced by using cuts on hadronic activity or missing energy. Hadronic activity is characterized by the variable $H_T$, the scalar sum of the transverse jet energies for all jets passing the preselection cuts. The missing transverse energy (MET) is the magnitude of the vector sum of the momenta of all candidate particles. Both $H_T$ and MET are sensitive discriminating observables for new physics, including the Higgs.

The background reduction ability of $H_T$ and MET are exploited in the following manner: Events with $H_T > 200$ (MET $>50$) GeV are said to have ``high'' $H_T$ (MET), while those with $H_T < 200$ (MET $< 50$) GeV are said to have ``low'' $H_T$ (MET). The high $H_T$ and high MET requirements (individually or in combination) lead to a significant reduction in Standard Model backgrounds. It is also possible to reduce backgrounds using the $S_T$ variable (the scalar sum of MET, $H_T$, and leptonic $p_T$), but for simplicity -- and to most closely mirror the analysis of  \cite{CMSMulti} -- we will not make strong use of $S_T$ here.

Further background reduction may be accomplished with a ``$Z$ veto'', in which the invariant mass of opposite-sign same-flavor (OSSF) lepton pairs is required to lie outside a $75-105$ GeV window around the $Z$ mass; we simply denote events passing the $Z$ veto as ``no $Z$''. In the case of $3 \ell$ events, it is also useful to differentiate between events with no OSSF pairs (which we label ``DY0'', i.e., no possible Drell-Yan pairs) and one OSSF pair (DY1). Although the current CMS multi-lepton analysis also includes channels with one or more hadronic $\tau$, in this analysis we will focus our attention on $\ell = e^\pm, \mu^\pm$ only. We include leptonic $\tau$'s in our analysis, classifying them according to their leptonic final state.

We note that there is also considerable sensitivity to new physics in the same-sign (SS) dilepton channel, although a SS dilepton channel is not currently included in the CMS multi-lepton analysis. We consider same-sign dileptons using the analysis cuts of the most recent CMS publication at 35 pb$^{-1}$ \cite{CMSdilepton}. For simplicity we include a SS dilepton search channel that places the most stringent constraints on the Higgs,  which in this case corresponds to requiring the $p_T$ of the leading lepton  $> 20$ GeV; the $p_T$ of both SS leptons $> 10$ GeV;  $H_T > 80$ GeV; and MET $>100$ GeV.

Dividing the multi-lepton signals into $3 \ell$ or $4 \ell$ events, there are 20 possible combinations of $H_T$ high/low; MET high/low; $Z$/no $Z$; and DY0/DY1; these may be aggregated to form the 11 channels used in \cite{CMSMulti}, with the addition of one further channel for SS dileptons. The collected channels  are presented in Table~\ref{tab:SM}. For each of the $3 \ell$ and $4 \ell$ categories, channels are presented in approximately descending order of sensitivity, with the last such channel dominated by SM backgrounds.

\begin{table}
\begin{center}
{\small
\begin{tabular}{lllcccccc}
\hline \hline \\
    & &  &   Expected & \multicolumn{4}{c}{Standard Model Higgs Boson} \\
& & &  Background &  120 & 130 & 140  & 150   \\
& & &   &  GeV       &   GeV      &   GeV     &    GeV   \\
   4 Leptons & \\ \\
$\dagger$\,MET ALL & HT HIGH &  ~~~~~  & $0 \pm 1.4 $ & 0.05 & 0.12 & 0.13 & 0.16  \\ 
$\dagger$\,MET HIGH & HT LOW & ~~~~~   & $ 0.57 \pm 0.32$ & 0.25 & 0.43 & 0.52 & 0.68 \\ 
$\dagger$\,MET LOW & HT LOW & No Z     & $0.53 \pm 0.33$ & 0.42 & 0.59 & 0.73 & 0.61 \\ 
& & & & (0.154) & (0.280) & (0.452) & (0.443) \\ 
$\dagger$\,MET LOW & HT LOW &~~~~~Z          & $30 \pm 3.4$ & 0.83 & 2.27 & 3.82 & 4.46\\ 
& & & & (0.351) & (0.598) & (0.787) & (0.950) \\

  & & \\
  3 Leptons &\\  \\

$\dagger$\,MET ALL & HT HIGH & DY0              & $0 \pm 1.2$ & 0.18 & 0.22 & 0.32 & 0.35  \\ 
$\dagger$\,MET ALL & HT LOW & DY0               & $34 \pm 5.4$ & 1.2 & 1.6 & 1.7 & 2.1 \\ 
~~MET ALL & HT HIGH & DY1~~~~~~   & $44 \pm 23$ & 0.60 & 1.1 & 1.4 & 1.6 \\ 
$\dagger$\,MET HIGH & HT LOW & DY1 No Z       & $26 \pm 9.9$ & 1.5 & 2.1 & 2.8 & 3.4 \\ 
~~MET HIGH & HT LOW & DY1~~~~~~Z    & $130 \pm 18$ & 1.3 & 1.7 & 2.1 & 2.3 \\ 
~~MET LOW & HT LOW & DY1 No Z          & $230 \pm 45$ & 1.5 & 2.5 & 3.0 & 3.0 \\ 
~~MET LOW & HT LOW & DY1~~~~~~Z    & $890 \pm 120$ & 3.2 & 5.1 & 6.3 & 6.7 \\ 

 & & \\
2 Leptons \\ \\
~~MET 100 & HT 80 & SS &  $50 \pm 20$ & 1.0 & 1.3 & 1.9 & 2.1  \\

   & &  \\
\hline \hline
\end{tabular}}
\caption{Expected number of Standard Model Higgs boson signal
and background events with systematic uncertainty
in various multi-lepton channels after acceptance and efficiency for 5 fb$^{-1}$ of 7 TeV proton-proton collisions. HIGH and LOW for MET and HT indicate $\MET \GTLT $ 50 GeV and $H_T \GTLT  200$ GeV
respectively. MET 100 and HT 80 indicate $\MET > 100$ GeV and $H_T > 80$ GeV respectively,
while ALL indicates no requirement on MET or HT.
DY0 $\equiv \ell^{\prime \pm}
\ell^{\mp} \ell^{\mp}$, DY1 $\equiv \ell^{\pm} \ell^+ \ell^-, \ell^{\prime \pm} \ell^+ \ell^-
$, and SS $\equiv \ell^{\pm} \ell^{\pm},\ell^{\pm} \ell^{\prime \pm}$, all for $\ell = e, \mu$.
No Z and Z indicate $|m_{\ell \ell} - m_Z| \GTLT 15$ GeV for any opposite sign same flavor
pair.
The numbers in parentheses
include only non-resonant four lepton contributions with $|m_{\ell \ell \ell \ell} - m_h| > 5$
GeV. The most sensitive channels are indicated with daggers.
}
\label{tab:SM}
\end{center}

\end{table}

Much of our analysis will focus on projected sensitivity at 5 fb$^{-1}$ of data. We estimate the Standard Model backgrounds (and corresponding systematic errors at CMS) for each of the $3\ell,4 \ell$ channels using the results presented in \cite{CMSMulti}. Specifically, we linearly extrapolate the background of  \cite{CMSMulti} from 35 pb$^{-1}$ to 5 fb$^{-1}$. For the systematic error on the background estimate, we extrapolate the error associated with Monte Carlo systematics linearly and the error associated with data-driven systematics as the square root of the integrated luminosity, as this should improve with an increasing sample size. For the SS dilepton channel, we make a rudimentary estimate of the background and systematic errors of \cite{CMSdilepton} for our choice of cuts, linearly extrapolated from 35 pb$^{-1}$ to 5 fb$^{-1}$. In the
$4 \ell$ [MET all, $H_T$ high] and $3 \ell$ [MET all, $H_T$ high, DY0] channels the backgrounds are expected to be small, but with the indicated uncertainties.

\subsection{Benchmark Higgs bosons}\label{subsec:channels}

Let us now turn to a consideration of the signal hypothesis. In order to evaluate the suitability of multi-lepton searches for the Higgs, we consider production and decays of the Standard Model Higgs and two simple variations: a ``Fermi-phobic'' Higgs with couplings only to SM gauge bosons; and a ``$b$-phobic'' Higgs with no couplings to down-type quarks ($d, s, b$)~\cite{fermiophobic,Gunion:1989we}. These two models are not meant to be taken literally but they represent idealized limits in the space of possible couplings, chosen for maximal clarity. Physical two-higgs doublet models with mass mixing and finite $v_1/v_2$ fully interpolate between the various coupling limits and, correspondingly, the population of various signal channels. Phenomenological properties of these scenarios, including production and decay modes, can be significantly different than the SM case; the most relevant aspects of this were explored in~~\cite{2hdmpheno} and will be emphasized in the paragraphs below. Multi-lepton signatures of intermediate models may be readily obtained from the simple limits presented here. Note that in both cases, we consider only the multi-lepton signals of the lightest neutral higgs boson in these models, having decoupled the remaining scalars.

In order to explore the full range of signals in the low-mass region, we focus on four representative mass points for detailed study: 120, 130, 140, and 150 GeV. The first three mass points bracket the currently-allowed low-mass region for the Standard Model Higgs. The final mass point is included for completeness since, e.g., a Fermi-phobic Higgs at 150 GeV is not yet excluded.

The multi-lepton search is most sensitive to Higgs bosons with appreciable branching ratios to $W$ and $Z$ gauge bosons. While the Standard Model Higgs decays predominantly to $WW^*, ZZ^*$ for $m_h \gtrsim 140$ GeV, at lower masses SM Higgs decays are dominated by lepton-free channels such as $h \to b \bar b$. Thus we are motivated to consider simple variations with enhanced branching ratios to gauge bosons, for which the multi-lepton search may offer maximal sensitivity. In this respect the Fermi-phobic and $b$-phobic models are not intended to exhaust the entirety of possible models, but rather represent idealized limits that bracket the multi-lepton signature space.

The Fermi-phobic Higgs is distinguished by the lack of any coupling to Standard Model fermions. Such a Higgs boson arises naturally in, e.g., a two-Higgs doublet model (2HDM) where only the second doublet couples to SM fermions and there is no mass mixing between the two doublets. In the $v_1 / v_2 \to \infty$ limit the tree-level couplings of the lightest (and Fermi-phobic) Higgs boson to $W$ and $Z$ bosons are the same as in the Standard Model. Although the absence of top loops reduces the one-loop effective $h \gamma \gamma$ coupling, appreciable contributions from $W$ loops remain. In contrast, the one-loop $h g g$ coupling vanishes. Hence the primary production channels for the Fermi-phobic Higgs are VBF and Higgs-strahlung, while decays are dominated by $h \to WW^*, \, ZZ^*, \, \gamma \gamma,$ and $Z \gamma$.

The $b$-phobic Higgs is characterized by the lack of any coupling to the down-type Standard Model quarks $d, s,$ and $b$. Tree-level couplings to gauge bosons, up-type quarks ($u, c, t$), and leptons are the same as in the Standard Model. Such couplings arise in a 2HDM where the leptons and $Q=2/3$ right-handed quarks couple to one doublet, while the $Q= -1/3$ right-handed quarks couple to the other doublet. In the limit where the two doublets are exact mass eigenstates and $v_1 / v_2 \to \infty$, the lightest (and $b$-phobic) Higgs boson couples to gauge bosons, leptons, and up-type quarks with Standard Model strength as desired. In this case, the one-loop effective $h \gamma \gamma$ and $h gg$ couplings are the same as in the Standard Model. Production channels of the $b$-phobic Higgs are the same as in the Standard Model, while decays are dominated by $h \to WW^*, gg, \tau^+ \tau^-$ at low masses and $h \to WW^*, ZZ^*$ at higher masses.

The production rates and branching ratios of the Standard Model Higgs are fixed by SM gauge couplings and fermion masses. The cross sections for each SM Higgs boson production channel and branching ratios for Higgs decays at each mass point were taken from the LHC Higgs Cross Section Group \cite{LHCHiggsCrossSectionWorkingGroup:2011ti}. The $gg$-fusion cross section is computed to NNLO$_{QCD}$ + NNLL$_{QCD}$ + NLO$_{EWK}$ precision; the weak VBF and associated $WH, ZH$ cross sections are computed to NNLO$_{QCD}$ + NLO$_{EWK}$ precision; and $t \bar t H$ is computed to NLO$_{QCD}$ precision. Cross sections and branching ratios for the Fermi-phobic Higgs were likewise taken from the LHC Higgs Cross Section Group \cite{LHCHiggsCrossSectionWorkingGroup:2011ti}. Cross sections and branching ratios for the $b$-phobic Higgs were obtained from the corresponding Standard Model values by setting the branching ratios $BR(h \to b \bar b) = BR(h \to s \bar s) = 0.$

For simulating signal processes, we have used MadGraph v4 \cite{Maltoni:2002qb,Alwall:2007st} and rescaled the cross sections to match the NLO results described above. For the production channels of $Wh$, $Zh$, $q\bar{q}h$ and $t\bar{t}h$, the Higgs boson was decayed in the $WW^*$, $ZZ^*$, $\tau^{+}\tau^{-}$ modes using BRIDGE \cite{Meade:2007js}. For the $gg\rightarrow h$ channel, the parton-level generation was done entirely within MadGraph with four charged leptons in the final state, thus including the effects of both $Z$'s going off-shell, as well as the contribution from two on-shell $Z$'s with the Higgs boson being off-shell. For the gluon fusion channel, the Higgs width was taken in accordance with  \cite{LHCHiggsCrossSectionWorkingGroup:2011ti}. Subsequent showering and hadronization effects were simulated using Pythia \cite{Sjostrand:2006za}. Detector effects were simulated using PGS \cite{PGS} with the isolation algorithm for muons and taus modified to more accurately reflect the procedure used by the CMS collaboration. In particular, we introduce a new output variable called ``\texttt{trkiso}'' for each muon or tau. The variable \texttt{trkiso} is defined to be the sum $p_T$ of all tracks, ECAL, and HCAL deposits within an annulus of inner radius 0.03 and outer radius 0.3 in $\Delta R$ surrounding a given muon or tau. Isolation requires that for each muon or tau, \texttt{trkiso}/$p_T$ of the muon or tau be less than 0.15. The efficiencies of PGS detector effects were normalized by simulating the mSUGRA benchmark studied in \cite{CMSMulti} and comparing the signal in 3$\ell$ and $4 \ell$ channels. To match efficiencies with the CMS study we applied an efficiency correction of 0.87 per lepton to our signal events. As discussed earlier, we applied preselection and analysis cuts  in accordance with those in \cite{CMSMulti}. 

\subsection{Multi-lepton signals of the Higgs}

The results of our signal simulation are presented in Tables~\ref{tab:SM} - \ref{tab:BP}. Table~\ref{tab:SM} contains the Standard Model background and expected signal events for a Standard Model Higgs boson at 5 fb$^{-1}$ for $m_h = 120, 130, 140,$ and $150$ GeV, broken down into the channels discussed earlier. Tables~\ref{tab:FP} and \ref{tab:BP} contain the same results for a Fermi-phobic and $b$-phobic Higgs boson, respectively. In each table, we have marked with a ($\dagger $) the channels that provide the most stringent limits on Standard Model Higgs production. Which channels provide the best limits is a sensitive balance of both signal and background.

\begin{table}
\begin{center}
{\small
\begin{tabular}{lllcccccc}
\hline \hline \\
    & &  &   Expected & \multicolumn{4}{c}{Fermi-phobic Higgs Boson} \\
& & &  Background &  120 & 130 & 140  & 150   \\
& & &   &  GeV       &   GeV      &   GeV     &    GeV   \\
   4 Leptons & \\ \\
$\dagger$\,MET ALL & HT HIGH &  ~~~~~ & $0 \pm 1.4$ & 0.04 & 0.10 & 0.07 & 0.06   \\ 
$\dagger$\,MET HIGH & HT LOW & ~~~~~   & $ 0.57 \pm 0.32$ & 0.65 & 0.81 & 0.58 & 0.64 \\ 
$\dagger$\,MET LOW & HT LOW & No Z     & $0.53 \pm 0.33$ & 0.23 & 0.15 & 0.19 & 0.12 \\ 
& & & & (0.23) & (0.15) & (0.19) & (0.12) \\ 
$\dagger$\,MET LOW & HT LOW &~~~~~Z          & $30 \pm 3.4$ & 0.94 & 1.22 & 0.97 & 0.86 \\ 
& & & & (0.70) & (0.80) & (0.54) & (0.49) \\

  & & \\
  3 Leptons &\\  \\

$\dagger$\,MET ALL & HT HIGH & DY0              & $0 \pm 1.2$ & 0.00 & 0.00 & 0.03 & 0.02 \\ 
$\dagger$\,MET ALL & HT LOW & DY0               & $34 \pm 5.4$ & 3.87 & 3.07 & 2.16 & 2.15 \\ 
~~MET ALL & HT HIGH & DY1~~~~~~   & $44 \pm 23$ & 0.32 & 0.68 & 0.66 & 0.53  \\ 
$\dagger$\,MET HIGH & HT LOW & DY1 No Z       & $26 \pm 9.9$ & 4.8 & 4.0 & 3.6 & 3.4 \\ 
~~MET HIGH & HT LOW & DY1~~~~~~Z    & $130 \pm 18$ & 4.1 & 3.3 & 2.8 & 2.6 \\ 
~~MET LOW & HT LOW & DY1 No Z          & $230 \pm 45$ & 3.5 & 3.1 & 2.3 & 2.0 \\ 
~~MET LOW & HT LOW & DY1~~~~~~Z    & $890 \pm 120$ & 9.5 & 7.4 & 5.5 & 4.6 \\ 

 & & \\
2 Leptons \\ \\
~~MET 100 & HT 80 & SS &  $50 \pm 20$ & 1.1 & 0.65 & 0.94 & 0.86 \\

   & &  \\
\hline \hline
\end{tabular}}
\caption{Expected number of Fermi-phobic Higgs boson signal
and background events with systematic uncertainty
in various multi-lepton channels after acceptance and efficiency for 5 fb$^{-1}$ of 7 TeV proton-proton collisions. HIGH and LOW for MET and HT indicate $\MET \GTLT $ 50 GeV and $H_T \GTLT  200$ GeV
respectively. MET 100 and HT 80 indicate $\MET > 100$ GeV and $H_T > 80$ GeV respectively,
while ALL indicates no requirement on MET or HT.
DY0 $\equiv \ell^{\prime \pm}
\ell^{\mp} \ell^{\mp}$, DY1 $\equiv \ell^{\pm} \ell^+ \ell^-, \ell^{\prime \pm} \ell^+ \ell^-
$, and SS $\equiv \ell^{\pm} \ell^{\pm},\ell^{\pm} \ell^{\prime \pm}$, all for $\ell = e, \mu$.
No Z and Z indicate $|m_{\ell \ell} - m_Z| \GTLT 15$ GeV for any opposite sign same flavor
pair.
The numbers in parentheses
include only non-resonant four lepton contributions with $|m_{\ell \ell \ell \ell} - m_h| > 5$
GeV. The most sensitive channels are indicated with daggers.
}
\label{tab:FP}
\end{center}

\end{table}

\begin{table}
\begin{center}
{\small
\begin{tabular}{lllcccccc}
\hline \hline \\
    & &  &   Expected & \multicolumn{4}{c}{$b$-phobic Higgs Boson} \\
& & &  Background &  120 & 130 & 140  & 150   \\
& & &   &  GeV       &   GeV      &   GeV     &    GeV   \\
   4 Leptons & \\ \\
$\dagger$\,MET ALL & HT HIGH &  ~~~~~  & $0 \pm 1.4$ & 0.14 & 0.25 & 0.19 & 0.19    \\ 
$\dagger$\,MET HIGH & HT LOW & ~~~~~   & $ 0.57 \pm 0.32$ & 0.70 & 0.85 & 0.76 & 0.81 \\ 
$\dagger$\,MET LOW & HT LOW & No Z     & $0.53 \pm 0.33$ & 1.2 & 1.2 & 1.1 & 0.73\\ 
& & & & (0.44) & (0.56) & (0.66) & (0.53) \\ 
$\dagger$\,MET LOW & HT LOW &~~~~~Z          & $30 \pm 3.4$ & 2.4 & 4.5 & 5.6 & 5.3 \\ 
& & & & (1.0) & (1.2) & (1.2) & (1.1) \\

  & & \\
  3 Leptons &\\  \\

$\dagger$\,MET ALL & HT HIGH & DY0              & $0 \pm 1.2$ & 0.50 & 0.44 & 0.47 & 0.42 \\ 
$\dagger$\,MET ALL & HT LOW & DY0               & $34 \pm 5.4$ & 3.5 & 3.2 & 2.5 & 2.5  \\ 
~~MET ALL & HT HIGH & DY1~~~~~~   & $44 \pm 23$ & 1.7 & 2.1 & 2.0 & 1.9\\ 
$\dagger$\,MET HIGH & HT LOW & DY1 No Z       & $26 \pm 9.9$ & 4.3 & 4.2 & 4.1 & 4.0 \\ 
~~MET HIGH & HT LOW & DY1~~~~~~Z    & $130 \pm 18$ &  3.7 & 3.3 & 3.0 & 2.8 \\ 
~~MET LOW & HT LOW & DY1 No Z          & $230 \pm 45$ & 4.3 & 5.0 & 4.4 & 3.6  \\ 
~~MET LOW & HT LOW & DY1~~~~~~Z    & $890 \pm 120$ & 8.9 & 9.9 & 9.2 & 7.9 \\ 

 & & \\
2 Leptons \\ \\
~~MET 100 & HT 80 & SS &  $50 \pm 20$ &  3.0 & 2.6 & 2.7 & 2.5 \\

   & &  \\
\hline \hline
\end{tabular}}
\caption{Expected number of $b$-phobic Higgs boson signal
and background events with systematic uncertainty
in various multi-lepton channels after acceptance and efficiency for 5 fb$^{-1}$ of 7 TeV proton-proton collisions. HIGH and LOW for MET and HT indicate $\MET \GTLT $ 50 GeV and $H_T \GTLT  200$ GeV
respectively. MET 100 and HT 80 indicate $\MET > 100$ GeV and $H_T > 80$ GeV respectively,
while ALL indicates no requirement on MET or HT.
DY0 $\equiv \ell^{\prime \pm}
\ell^{\mp} \ell^{\mp}$, DY1 $\equiv \ell^{\pm} \ell^+ \ell^-, \ell^{\prime \pm} \ell^+ \ell^-
$, and SS $\equiv \ell^{\pm} \ell^{\pm},\ell^{\pm} \ell^{\prime \pm}$, all for $\ell = e, \mu$.
No Z and Z indicate $|m_{\ell \ell} - m_Z| \GTLT 15$ GeV for any opposite sign same flavor
pair.
The numbers in parentheses
include only non-resonant four lepton contributions with $|m_{\ell \ell \ell \ell} - m_h| > 5$
GeV. The most sensitive channels are indicated with daggers.
}
\label{tab:BP}
\end{center}

\end{table}

In the $4 \ell$ final states, the strongest constraints on a Standard Model Higgs arise from the [MET high, $H_T$ low] channel  and from the two [MET low, $H_T$ low] channels, both with $Z$ and without $Z$. That these latter two channels provide a constraint on Higgs production is not surprising; here the signal is dominated by $h \to ZZ^* \to 4 \ell$ (where the $h$ is produced via $gg$F or VBF) and the same channels are used in the conventional gold-plated resonant $h \to 4 \ell$ Higgs search. However, it bears emphasizing that this existing search is sensitive only to resonant production of the four-lepton final state, while in fact these two channels are populated both by resonant {\it and} non-resonant signal events. Here ``resonant'' is taken to mean that $m_{4 \ell}$ is within $\pm 5$ GeV of the Higgs mass. In contrast, non-resonant events receive significant contributions from, e.g., $h \to ZZ^* \to 2 \ell 2 \tau$.  For $m_h = 140$ GeV, the ratio of non-resonant to resonant events in the [MET low, $H_T$ low, $Z$] channel is small, of the order 1:4. But in the (low background) [MET low, $H_T$ low, no $Z$] channel the ratio is far larger, of the order 5:3! Thus a multi-lepton search that is sensitive to both non-resonant {\it and} resonant $4 \ell$ production may enjoy considerable advantages over conventional Higgs searches, insofar as non-resonant $h \to 4 \ell$ production contributes significantly to channels with low SM background.

In contrast, the remaining sensitive $4 \ell$ channel, [MET high, $H_T$ low], is dominated by an entirely different process, $Zh$ associated production with $h \to WW^*$.  At low masses the branching fraction $h \to WW^*$ drops off, but is largely compensated for by a rise in $h \to \tau^+ \tau^-$ with both $\tau$'s decaying leptonically. Additional contributions to the Higgs signal in this channel come from $t \bar t h$ associated production with $h \to WW^*$; these events tend to have high $H_T$, but a significant fraction fall below the $H_T$ cut.

In the $3 \ell$ final states, the most stringent limits come from [MET all, $H_T$ low, DY0]; [MET all, $H_T$ high, DY0]; and  [MET high, $H_T$ low, DY1 no $Z$]. For both low-$H_T$ channels the primary contribution to signal comes from associated $Wh$ production with $h \to WW^*$. However, as with the $4 \ell$ events, an additional contribution arises from associated $t \bar t h$ production with $h \to WW^*$ that falls below the $H_T$ cut. Likewise, the  [MET all, $H_T$ high, DY0] channel is dominated by $t \bar t h$ with $h \to WW^*$.  In all cases, the decrease in $h \to WW^*$ at low masses is compensated by a rise in $h \to \tau^+ \tau^-$.

\begin{figure}[htbp]
\begin{center}
\includegraphics[scale=0.7]{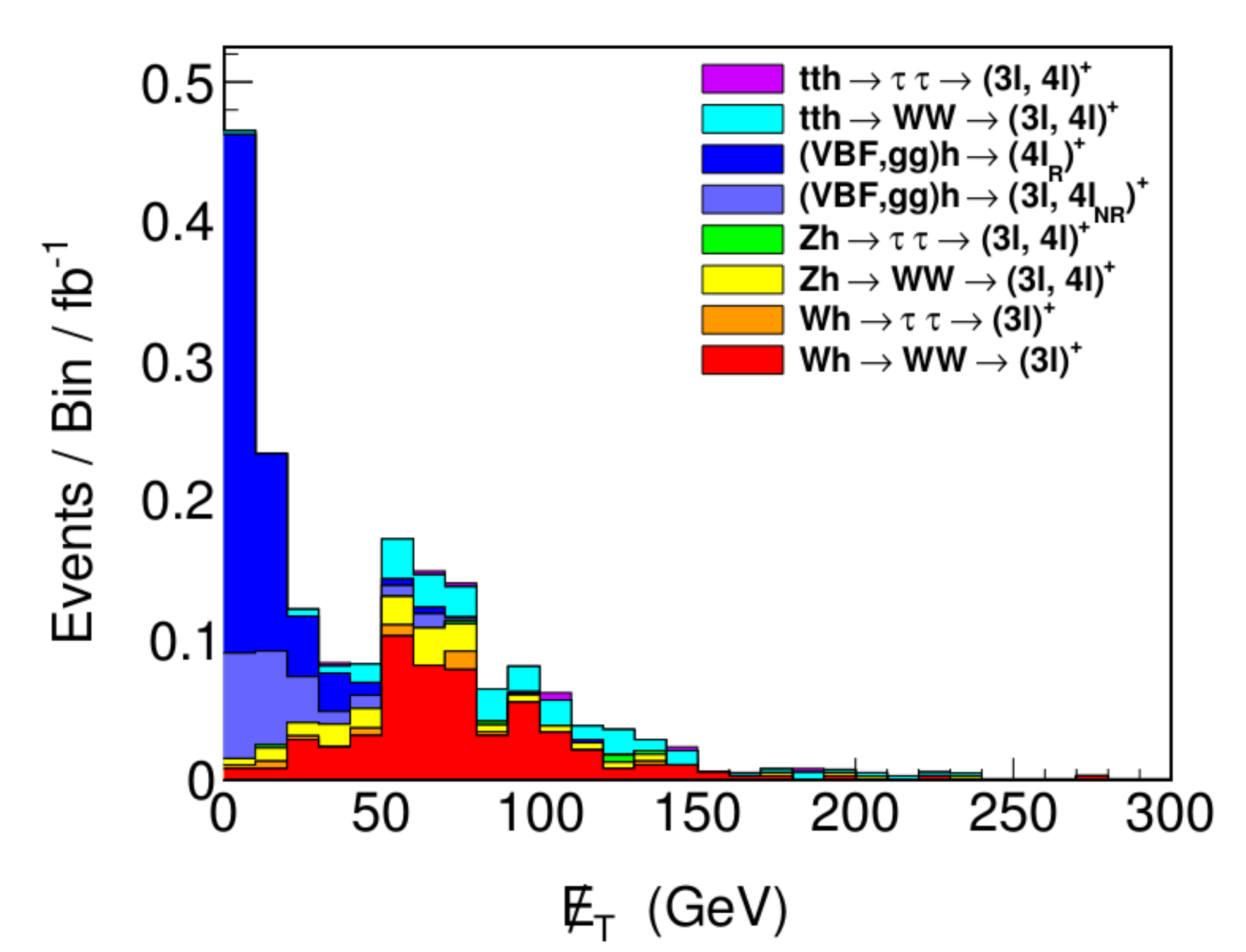}
\caption{The missing transverse energy distribution
after acceptance and efficiency
for 7 TeV proton-proton collisions
arising from
Standard Model 140 GeV Higgs boson contributions to the
three and four lepton channels indicated with daggers in Table 1.
Individual Higgs boson contributions are shown for
production in association with $W$ and $Z$ vector bosons and $t \bar{t}$ top quark
pairs with $h \to WW^*, \tau \tau$
including all $W, Z$ and $\tau$ decay modes,
as well as
production from gluon-gluon and vector boson fusion
with $h \to Z^*Z^{(*)}$ with all
$Z$ decay modes that give
both resonant and non-resonant contributions to the
four lepton dagger channels including through $\tau \tau$.
The bin size is 10 GeV and the highest bin includes overflow.}
\label{fig:met140}
\end{center}
\end{figure}

\begin{figure}[htbp]
\begin{center}
\includegraphics[scale=0.7]{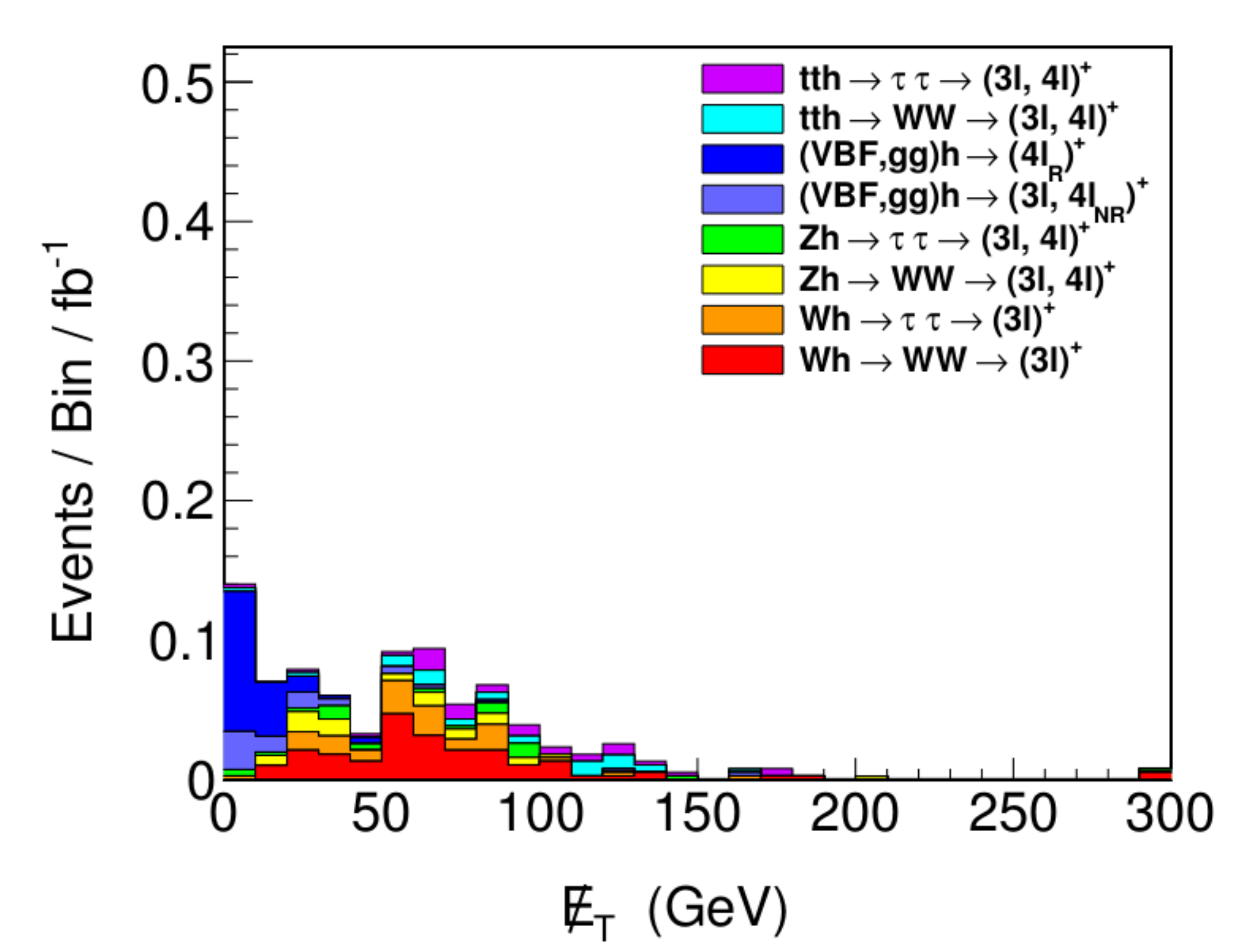}
\caption{The missing transverse energy distribution in the three and four lepton channels after accounting for acceptance and efficiency for a 120 GeV Standard Model Higgs Boson produced in 7 TeV proton-proton collisions. Individual Higgs boson contributions are shown for
production in association with $W$ and $Z$ vector bosons and $t \bar{t}$ top quark
pairs with $h \to WW^*, \tau \tau$
including all $W, Z$ and $\tau$ decay modes,
as well as
production from gluon-gluon and vector boson fusion
with $h \to Z^*Z^{(*)}$ with all
$Z$ decay modes that give
both resonant and non-resonant contributions to the
four lepton dagger channels including through $\tau \tau$. The bin size is 10 GeV and the highest bin includes overflow.}
\label{fig:met120}
\end{center}
\end{figure}

\begin{figure}[htbp]
\begin{center}
\includegraphics[scale=0.7]{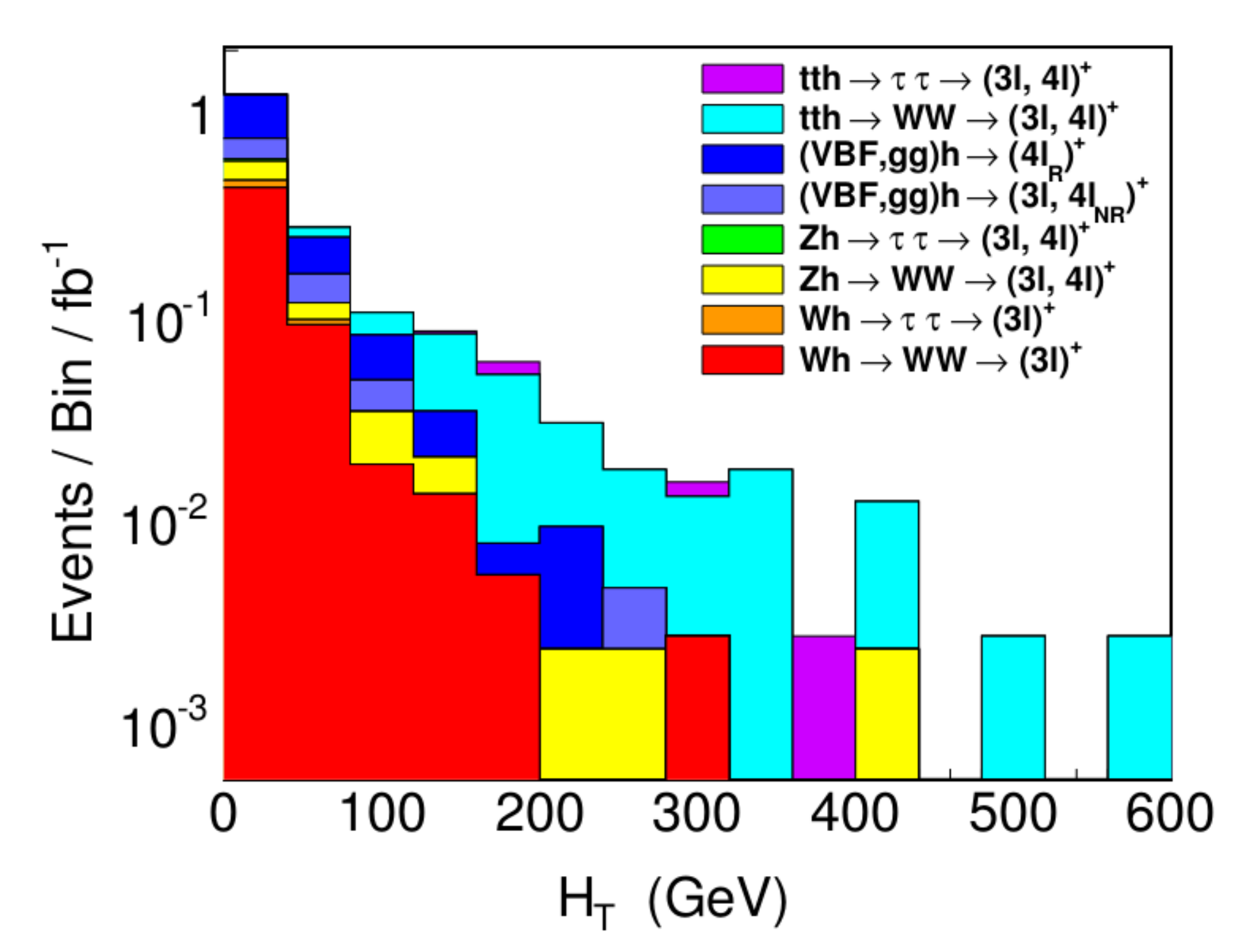}
\caption{The transverse hadronic energy distribution summed over jets with $p_T > 40$ GeV after accounting for acceptance and efficiency for a 140 GeV Standard Model Higgs Boson produced in 7 TeV proton-proton collisions. Individual Higgs boson contributions are shown for
production in association with $W$ and $Z$ vector bosons and $t \bar{t}$ top quark
pairs with $h \to WW^*, \tau \tau$
including all $W, Z$ and $\tau$ decay modes,
as well as
production from gluon-gluon and vector boson fusion
with $h \to Z^*Z^{(*)}$ with all
$Z$ decay modes that give
both resonant and non-resonant contributions to the
four lepton dagger channels including through $\tau \tau$. The bin size is 40 GeV and the highest bin includes overflow.}
\label{fig:ht140}
\end{center}
\end{figure}

\begin{figure}[htbp]
\begin{center}
\includegraphics[scale=0.7]{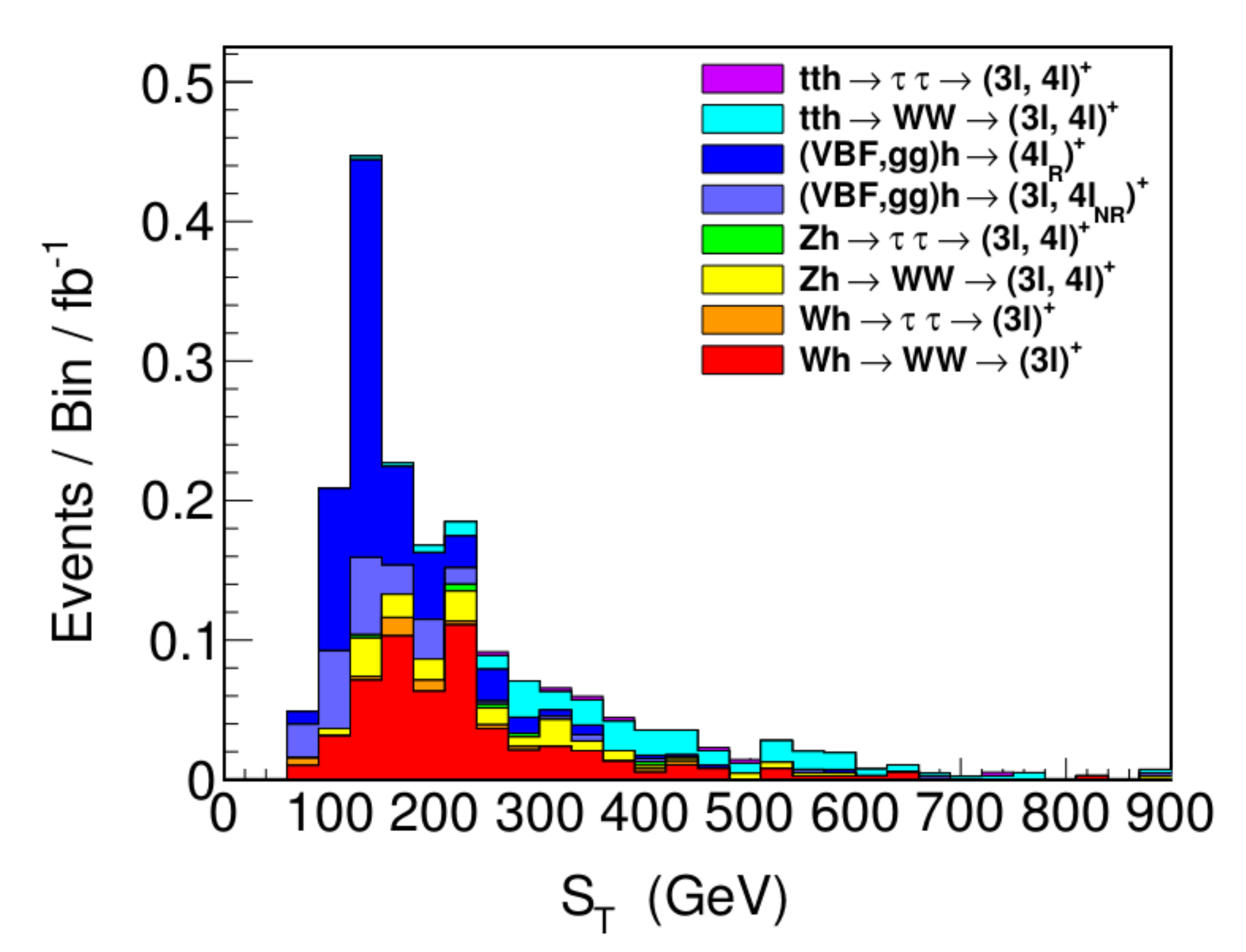}
\caption{The total transverse energy distribution summed over leptons, missing energy, and jets with $p_T > 40$ GeV after accounting for acceptance and efficiency for a 140 GeV Standard Model Higgs Boson produced in 7 TeV proton-proton collisions. Individual Higgs boson contributions are shown for
production in association with $W$ and $Z$ vector bosons and $t \bar{t}$ top quark
pairs with $h \to WW^*, \tau \tau$
including all $W, Z$ and $\tau$ decay modes,
as well as
production from gluon-gluon and vector boson fusion
with $h \to Z^*Z^{(*)}$ with all
$Z$ decay modes that give
both resonant and non-resonant contributions to the
four lepton dagger channels including through $\tau \tau$. The bin size is 30 GeV and the highest bin includes overflow.}
\label{fig:st140}
\end{center}
\end{figure}

These contributions may all be seen visually in the MET, $H_T$, and $S_T$ distributions of $3 \ell$ and $4 \ell$ Higgs signal events, as illustrated in figures \ref{fig:met140}-\ref{fig:st140}. Figure~\ref{fig:met140} shows the MET distribution for a Standard Model Higgs boson at 140 GeV. Figure~\ref{fig:met120} shows the MET distribution for a lighter Standard Model Higgs at 120 GeV. Figures~\ref{fig:ht140} and \ref{fig:st140} show the $H_T$ and $S_T$ distributions for a 140 GeV Standard Model Higgs, respectively. In each figure only the most important production channels are plotted: $Wh$, $Zh$, and $t \bar t h$ associated production with $h \to WW^*, \tau^+ \tau^-$; and $gg$F, VBF with $h \to ZZ^* \to 4l$. Similarly, only the signal events populating the constraining signal channels (the channels marked with daggers in the tables) are shown. Several features are salient. While Higgs production does not lead to exceptionally large MET production, the MET distribution is peaked far enough out to populate high MET channels with low Standard Model backgrounds. The hadronic activity of resonant and $Wh$, $Zh$ associated production is small, fully populating the low-$H_T$ channels, while $t \bar t h$ effectively populates high-$H_T$ and contributes close to the $H_T$ cut. Much as with MET, the $S_T$ of Higgs events is not exceptionally large, but peaks at sufficiently high values to differentiate from Standard Model backgrounds. Finally, the contribution of $h \to \tau^+ \tau^-$ to signal events at low masses is readily apparent in the MET plot for $m_h = 120$ GeV.


The story is quite similar for both variant Higgs models. In the case of the Fermi-phobic Higgs, the same $4 \ell$ channels provide maximum sensitivity. Indeed, the enhanced branching rate for $h \to WW^*$ results in particularly strong constraints in the [MET high, $H_T$ low] channel from $Zh$ associated production, though this is partially compensated for by the disappearance of $t \bar t h$ associated production. In the $3 \ell$ channels, the limits from both constraining channels strengthen relative to the SM due to the enhanced $h \to WW^*$ rate. While existing $h \to \gamma \gamma$ searches provide the greatest sensitivity to the Fermi-phobic Higgs at very low masses ($\lesssim 120$ GeV), the considerable sensitivity to $Wh, Zh$ associated production with $h \to W W^*$ may make a Higgs multi-lepton search the most powerful approach at higher masses.

For the $b$-phobic Higgs, the disappearance of $h \to b \bar b$ simply increases the production rate in multi-lepton signal channels relative to the Standard Model, particularly at low Higgs masses. The same $3 \ell, 4 \ell$ channels remain the most constraining, albeit with enhanced sensitivity.

Although limits may be placed on Higgs production due to any individual channel in the multi-lepton search, the greatest sensitivity comes from combining all channels. In Table~\ref{tab:CL} we show the expected 95\% CL limits on the production cross section of our benchmark models as a multiple of the corresponding theory cross section. These limits represent a combined Bayesian 95\% CL limit computed at 5 fb$^{-1}$ using the background estimates and systematic errors listed in Table~\ref{tab:SM} with $n_{observed} = n_{expected}$, i.e., an expected limit computed without considering the results of the 35 pb$^{-1}$ study  \cite{CMSMulti}. We differentiate between the limits set by all contributions (including the resonant $h \to 4 \ell$ final states present in the existing golden channel search) and those set by purely non-resonant contributions unique to the multi-lepton search. Notably, the current search strategy may already limit cross sections on the order of a few times the Standard Model value. Equally noteworthy is the sensitivity to a Fermi-phobic Higgs at high masses, $m_h \gtrsim 130$ GeV. Current searches for a Fermi-phobic Higgs are driven primarily by the enhanced rate for $h \to \gamma \gamma$, and thus are primarily sensitive to low masses. 
Thus a dedicated multi-lepton search would provide some of the best limits on a Fermi-phobic Higgs above 130 GeV.

\begin{table}
\begin{center}
{\small
\begin{tabular}{lcccc}
\hline \hline \\
& & \\
& 120 GeV & 130 GeV & 140 GeV & 150 GeV \\
& & \\
All Contributions & & & & \\
~Standard Model Higgs & 4.3  & 2.7  & 2.0  & 1.8  \\
~Fermi-phobic Higgs & 2.2  & 2.3 & 2.9  & 3.0  \\
~$b$-phobic Higgs & 1.6 & 1.4  & 1.4 & 1.5 \\
& & \\
Non-resonant Contributions & & & & \\
~Standard Model Higgs & 5.8 & 3.8  & 3.0  & 2.6  \\
~Fermi-phobic Higgs & 2.2 & 2.4 & 3.1 & 3.2  \\
~$b$-phobic Higgs & 2.0 & 2.0  & 2.1  & 2.2  \\
  & &  \\
\hline \hline
\end{tabular}}
\caption{Expected 95\% CL limits on the production cross section times branching ratio in multiples of the theory cross section times branching ratio for 5 fb$^{-1}$ of 7 TeV proton-proton collisions. Limits are obtained from an exclusive combination of all the multi-lepton channels in Tables 1$-$3 assuming the observed numbers of events are equal to the expected. The rows labelled All Contributions include contributions from Higgs boson production in association with $W$ and $Z$ vector bosons and $t \bar{t}$ top quark pairs with $h \to WW^*, ZZ^*, \tau^+ \tau^-$ including all $W, Z$ and $\tau$ decay modes, as well as production from gluon-gluon and vector boson fusion with $h \to Z^*Z^{(*)}$ with all $Z$ decay modes that give four leptons including through intermediate $\tau^+ \tau^-$. The rows labelled Non-resonant Contributions include the additional requirement that $|m_{\ell \ell \ell \ell} - m_h| > 5$ GeV for the four lepton channels.}
\label{tab:CL}
\end{center}
\end{table}

We emphasize that although the multi-lepton search may not appear to be as sensitive to the Standard Model Higgs boson as some of the current search strategies being pursued at ATLAS and CMS, the sensitivity shown in Table~\ref{tab:CL} corresponds to the current multi-lepton search strategy without any further optimization for a Higgs search. Sensitivity may readily be improved by further tailoring cuts, as we will discuss in \S\ref{sec:future}.


\section{Towards a dedicated multi-lepton Higgs search}\label{sec:future}

We have seen that the existing CMS multi-lepton search strategy has considerable sensitivity to the Standard Model Higgs and its variants, with the potential to exclude production cross sections of a few times the Standard Model value in the light mass window {\it without} specific tailoring to the Higgs signal. However, significant improvements in sensitivity may be achieved by refining the search strategy for a dedicated Higgs multi-lepton combination.

Among the $4 \ell$ channels with highest sensitivity to the Higgs, the [MET low, $H_T$ low, $Z$ / no $Z$] channels are already fairly optimized for the Higgs; they receive principal contributions from the $h \to ZZ^* \to 4 \ell$ golden mode as well as the  $h \to Z^*Z^* \to 4 \ell$ mode. The same is true of the [MET high, $H_T$ low] channel dominated by $Zh$ associated production with $h \to WW^*$. We emphasize, though, that the sensitivity of these channels to {\it nonresonant} $4 \ell$ production may give an appreciable advantage over the conventional golden mode search.

However, it is also important to emphasize the role of $t \bar t h$ associated production in potential $4 \ell$ signals. In particular, $t \bar t h$ associated production with $h \to WW^*$ contributes significantly to the $4 \ell$ [MET high, $H_T$ low] channel without the presence of a $Z$ boson. Dividing the $4 \ell$ [MET high, $H_T$ low] channel into two channels, with and without $Z$, would help to reduce backgrounds for this signal. Furthermore, these final states include two $b$ quarks from the decays of the tops. Since the primary background in this channel is from di-$Z$ production where one $Z$ is off-shell -- for which there are no $b$ quarks in the final state -- further discrimination may be obtained by requiring one or two $b$ tags in the final state. Requiring $b$-tags should also increase the sensitivity of other channels that receive a significant contribution from $t \bar t h$, particularly [MET all, $H_T$ high]. While this channel is not  the most sensitive of the $4 \ell$ channels, further reduction of the Standard Model background expectation -- perhaps by $b$-tags and the addition of a $Z$ veto -- may render it more useful. It should be emphasized, of course, that requiring $b$-tags in channels sensitive to $t \bar t h$ will not completely erase the Standard Model background expectation. In addition to di-$Z$ contributions to these channels, there may be considerable backgrounds from $t \bar t \gamma^*$ that are not accounted for in the current CMS search. These backgrounds would survive $b$ tag requirements, and should be carefully accounted for in a dedicated search.

Similar improvement may be attained by requiring one or two $b$-tags in the $3 \ell$ [MET all, $H_T$ high, DY0] and [MET all, $H_T$ high, DY1] channels, which are likewise dominated by $t \bar t h$. The remaining sensitive $3 \ell$ channels receive signals primarily from associated $Wh$ production with $h \to WW^*$, for which the existing cuts are adequately optimized.

Finally, we note that many of the sensitive search channels in both $3 \ell$ and $4 \ell$ final states receive significant contributions from $Wh, Zh$, and $t\bar t h$ associated production with $h \to \tau^+ \tau^-$, particularly for low Higgs masses ($m_h \lesssim 130$ GeV). These decays contribute directly to existing search channels when one or both of the $\tau$'s decay leptonically. However, some sensitivity is lost since the $\tau$ leptonic branching fraction is only $\sim 35\%$. Since it is possible to tag hadronically-decaying $\tau$'s with some degree of accuracy, sensitivity to associated production may be improved by adding channels for exclusive final states with, e.g., two leptons and one or two hadronic $\tau$'s.



\section{Conclusions}\label{sec:conc}

As the LHC experiments are analyzing the 5 fb$^{-1}$ data set, the potential for discovery or exclusion of the Standard Model Higgs boson grows ever greater. In this paper we have evaluated the possibility of augmenting existing LHC searches for the Higgs via the combination of channels with multiple non-resonant leptons. The total multi-lepton Higgs signal in these channels exceeds the gold-plated $4 \ell$ resonant mode, though it is spread over same-sign $2 \ell$, $3 \ell$, and $4 \ell$ final states. The exclusive combination of these channels using the existing CMS multi-lepton search strategy yields a sensitivity competitive with other discovery-level searches for the Higgs boson, both for the Standard Model Higgs and for variants with enhanced branching ratios to leptons and gauge bosons. Refinements focused specifically on the Higgs boson signal, such as $b$-tags in channels involving $t\bar t h$ associated production, would provide even more sensitivity. The extensive study of Standard Model backgrounds in current multi-lepton searches suggest that an effective multi-lepton search for the Higgs could be implemented fairly quickly.

Although we have focused in this paper on the sensitivity of a multi-lepton search for a single Higgs doublet, we emphasize that there may be even greater discovery potential for an extended Higgs sector with an enhanced total multi-lepton cross section, which we leave to a future study. The advantage of a multi-lepton search lies in its exclusive combination of multiple leptonic final states, both resonant and nonresonant alike. As such, it is sensitive to simultaneous contributions from more than one new state with appreciable leptonic decays. For example, in a two-Higgs doublet model the multi-lepton signals of the lightest neutral Higgs $h$ are augmented by new production and decay channels from the heavier neutral Higgs $H$, the pseudoscalar $A$, and the charged Higgses $H^\pm$. Processes such as $gg \to H \to hh \to WW^*WW^*$ and $gg \to A \to Zh \to ZWW^*$ contribute significantly to both $3 \ell$ and $4 \ell$ final states. Moreover, in such models the cross section for specific resonant final states such as $h \to \gamma \gamma$ and $h \to ZZ^* \to 4 \ell$ may be {\it suppressed} relative to the Standard Model expectation, reducing the effectiveness of existing resonant searches. Should the Higgs sector prove to be extended beyond a single electroweak doublet, a dedicated multi-lepton Higgs search may provide the most promising avenue for discovery.



\bigskip
\bigskip

{ \Large \bf Acknowledgments}

\smallskip \smallskip

\noindent
We thank Amit Lath and Matthew Walker for useful conversations. The research of NC, MP and ST was supported in part by DOE grant DE-FG02-96ER40959. The research of RG and SS was supported in part by NSF grant PHY-0969282. The research of CK was supported in part by NSF grant PHY-0969020. NC gratefully acknowledges the support of the Institute for Advanced Study.


\end{document}